\documentclass[11pt]{article} 
\usepackage{amsfonts}
\usepackage{amsmath}

\newtheorem{theorem}{Theorem}[section]
\newtheorem{lemma}[theorem]{Lemma}

\def\square{\rule{2mm}{2mm}}

\newenvironment{proof}{{\noindent\bf Proof:  }}{\qquad\square}

\def\squarebox#1{\hbox to #1{\hfill\vbox to #1{\vfill}}}

\newcommand{\tensor}{\otimes}

\newcommand{\xor}{\oplus}

\newcommand\meet\wedge

\newcommand{\complex}{{\mathbb C}}

\newcommand{\trace}{{\rm Tr}}

\newcommand{\ket}[1]{|#1\rangle}
\newcommand{\bra}[1]{\langle #1|}
\newcommand{\braket}[2]{\langle #1 | #2\rangle}
\newcommand{\ketbra}[2]{\ket{#1}\!\bra{#2}}
\newcommand{\density}[1]{\ketbra{#1}{#1}}
\newcommand{\norm}[1]{\left\|\,#1\,\right\|}
\newcommand{\trnorm}[1]{\norm{#1}_{\mathrm {tr}}}
\newcommand{\set}[1]{{\left\{#1\right\}}}

\newcommand{\ignore}[1]{}

\newcommand{\inr}{\in_{\mathrm R}}

\newcommand{\aitch}{{\mathcal H}}

\begin{document}

\title{\bf Weak coin flipping with small bias}

\author{
{\large Iordanis Kerenidis} \thanks{Supported by DARPA under agreement
  number F~30602-01-2-0524.}\\  
{\small Computer Science Division} \\
{\small University of California, Berkeley} \\
{\small jkeren@cs.berkeley.edu}
\and
{\large Ashwin Nayak} \thanks{Supported by Charles Lee Powell
  Foundation, and NSF grants CCR~0049092 and EIA~0086038.} \\
{\small Computer Science Department} \\
{\small California Institute of Technology} \\
{\small nayak@cs.caltech.edu}
}

\date{}

\maketitle

\begin{abstract}
This note presents a quantum protocol that demonstrates that {\em
  weak\/} coin flipping with bias~$\approx 0.239$, less than~$1/4$, is
possible. A bias of~$1/4$ was the smallest known, and followed from
the strong coin flipping protocol of Ambainis~\cite{Ambainis01}
(also proposed by Spekkens and Rudolph~\cite{SpekkensR02a}).
Protocols with yet smaller bias~$\approx 0.207$ have independently
been discovered~\cite{Ambainis01c,SpekkensR02b}. We also present an
alternative strong coin flipping protocol with bias~$1/4$ with
analysis simpler than that of~\cite{Ambainis01}. A similar analysis
for a class of cheating strategies has been given
by~\cite{SpekkensR02a}.
\end{abstract}
 
\section{Quantum weak coin flipping}

Often in applications based on this primitive, coin-flipping is used
to choose one of two competing parties as the ``winner''. In the
classic example from~\cite{Blum81}, Alice and Bob are getting a
divorce, and would like to decide who gets the car. They decide to
toss a coin for that purpose, but don't trust each other.  In such a
scenario, they could instead play any fair game to decide the issue.
Motivated by this, we consider the following weaker version of
coin-flipping.

A {\em weak coin flipping\/} protocol with bias~$\epsilon$, is a
two-party communication game in the style of~\cite{Yao93}, in which
the players start with no inputs, and compute a value~$c_A, c_B \in
\set{0,1}$ respectively or declare that the other player is
cheating. The protocol is deemed successful if Alice and Bob agree on
the outcome, i.e.~$c_A = c_B$. Then, the outcome~$0$ is identified
with Alice winning, and~$1$ with Bob winning. The protocol satisfies
the following additional properties:
\begin{enumerate}

\item If both players are honest (i.e., follow the protocol), then
  they agree on the outcome of the protocol: $c_A = c_B$, and the game
  is fair: $\Pr(c_A = c_B = b) = 1/2$, for $b \in \set{0,1}$.

\item If one of the players is honest (i.e., the other player may
  deviate arbitrarily from the protocol in his or her local
  computation), then the other party {\em wins\/} with probability
  at most~$1/2 + \epsilon$. In other words, if Bob is dishonest, then 
  $\Pr(c_A = c_B = 1) \le 1/2 + \epsilon$, and if Alice is dishonest,
  then $\Pr(c_A = c_B = 0) \le 1/2 + \epsilon$.

\end{enumerate}

In a strong coin flipping protocol, the goal is instead to produce a
random bit which is biased away from any particular value~$0$ or~$1$.
Clearly, any strong coin flipping protocol with bias~$\epsilon$
leads to weak coin flipping with the same bias. We may also derive a
strong coin-flipping protocol from a weak one. A simple way to do
this is to have the winner of the game flip the coin. This results in
an increase in the bias of the protocol, however: if when one player,
say Alice, is dishonest, and the other (Bob) honest, the probability
of Alice winning is~$p_w \ge 1/2$, and the probability of Bob winning
is~$p_\ell$, then the coin will have bias~$p_w + (p_\ell-1)/2$.

The primitive of quantum strong coin flipping has been studied in,
e.g.,~\cite{LC98,Mayers97,ATVY00,Ambainis01,SpekkensR02a}. The best
known protocol, with bias~$1/4 = 0.25$, is due to
Ambainis~\cite{Ambainis01}, also independently proposed by Spekkens
and Rudolph~\cite{SpekkensR02a}.  This note presents a protocol that
demonstrates that {\em weak\/} coin flipping with bias~$\approx
0.239$, less than~$1/4$, is possible. This protocol is obtained by
modifying the protocol of~\cite{Ambainis01} especially so that the
{\em winning\/} party is checked for cheating.  We also describe a
related strong coin flipping protocol with bias~$1/4$ that has the
advantage over~\cite{Ambainis01} that the analysis is considerably
simpler. A similar analysis for a class of cheating strategies has
been given by~\cite{SpekkensR02a}.

Since the discovery of the abovementioned protocol, we have learnt of
several exciting developments. Kitaev~\cite{Kitaev01} has shown that
in any protocol for {\em strong\/} coin flipping, the product of the
probabilities with which each of the players can achieve outcome (say)
$0$, has to be at least $1/2$. Hence the protocols with arbitrarily
small bias are not possible; the bias is always at
least~$1/\sqrt{2}-1/2 \approx 0.207$.  (Previous lower bounds applied
only to certain kinds of
protocol~\cite{Ambainis01,SpekkensR02a,NayakSh02}.)  Furthermore,
Ambainis~\cite{Ambainis01c} and Spekkens and
Rudolph~\cite{SpekkensR02b} have constructed a family of protocols for
{\em weak\/} coin flipping, where the product of the winning
probabilities is exactly~$1/2$. By making the winning probabilities
equal, they get protocols in which each player wins with probability
at most~$1/\sqrt{2}$, and hence the bias is~$1/\sqrt{2}-1/2\approx
0.207$. Subsequently, Ambainis~\cite{Ambainis01b} proved a lower bound
of~$1/2$ for the product of the winning probabilities for the specific
class of protocols considered in~\cite{SpekkensR02b}. We note that the
lower bound of Kitaev for strong coin flipping does not apply here and
hence quantum games of the weaker variety with even smaller bias may
be possible.

\section{A game with small bias}

Below, we describe a weak coin flipping game that has bias less
than~$1/4$. The game is derived from the protocol
of~\cite{Ambainis01}, which achieves the previously best known bias
of~$1/4$.

The protocol is parametrised by~$\alpha \in [0,\pi]$, which we will
optimise over later.  For~$x,s \in \set{0,1}$, define the
state~$\ket{\psi_{x,s}} = \ket{\psi_{x,s}(\alpha)}$ in a Hilbert space
${\mathcal H}_t= \complex^3$, and~$\ket{\psi_x} = \ket{\psi_x(\alpha)}
\in \aitch_s \tensor
\aitch_t = \complex^2 \tensor \complex^3$ as:
\begin{eqnarray}
\label{eqn-trit}
\ket{\psi_{x,s}}
    & = & \cos{\alpha\over 2} \ket{0} + (-1)^s \sin{\alpha\over 2}
          \ket{x+1} \\
\label{eqn-state}
\ket{\psi_x}
    & = & \frac{1}{\sqrt{2}}( \ket{0} \ket{\psi_{a,0}} + \ket{1}
          \ket{\psi_{a,1}}).
\end{eqnarray}
The protocol has the following rounds:
\begin{enumerate}
\item 
Alice picks~$a \inr \set{0,1}$, prepares the state~$\ket{\psi_{a}}$
in~$\aitch_s \tensor \aitch_t$ (i.e., over one qubit and one qutrit)
and sends Bob the right half of the state (the qutrit).

\item Bob picks~$b \inr \set{0,1}$ and sends it to Alice.
  
\item Alice then reveals the bit~$a$ to Bob. Let~$c = a \xor b$.
  
  If~$c = 0$, then~$c_A \leftarrow 0$ and she sends the other part of
  the state~$\ket{\psi_a}$ (the {\em sign\/} qubit).  Bob checks that
  the qutrit-qubit pair he received in the first and the current
  rounds are indeed in state~$\ket{\psi_a}$.  If the test is passed,
  Alice wins ($c_B \leftarrow 0$ as well), else Bob concludes that
  Alice has deviated from the protocol, and aborts.
  
\item 
If, on the other hand, $c = a \xor b = 1$, then~$c_B \leftarrow 1$,
and Bob returns the qutrit he received in round~$1$. Alice checks that
her qubit-qutrit pair are in state~$\ket{\psi_a}$. If the test is
passed, Bob wins the game ($c_A \leftarrow 0$), else, Alice concludes
that Bob has tampered with her qutrit to bias the game, and aborts.

\end{enumerate}

If the two players follow this protocol, the game is fair. We now
analyse the situation where one of the players cheats.

\begin{lemma}
\label{thm-alice}
If Bob is honest, then the probability that Alice wins~$\Pr(c_B = 0)
~\le~ {1 \over 2} (1 + \cos^2{\alpha\over 2})$.
\end{lemma}
\begin{proof} 
We assume w.l.o.g.\ that a dishonest Alice tries to maximize her
probability of winning, and therefore sends~$a = b$ (so that~$c = a
\xor b = 0$) in round~3. Her cheating strategy then takes the following form.
Alice uses some ancillary space ${\mathcal H}$ and prepares some state
$\ket{\psi}\in {\mathcal H}\otimes{\mathcal H}_s\otimes{\mathcal
H}_t$. She keeps the part of the state in ${\mathcal
H}\otimes{\mathcal H}_s$ and sends the qutrit part in ${\mathcal H}_t$
to Bob. Let~$\sigma$ denote the density matrix of Bob after the first
round of the protocol (i.e., of the qutrit). Let~$\rho_a$ be the
density matrix he would have if Alice had prepared the honest state
$\ket{\psi_a}$:
\[ 
\rho_a ~=~ \trace_{\aitch_a} \; \density{\psi_a}
       ~=~ {1\over 2}(\density{\psi_{a,0}}+\density{\psi_{a,1}})
       ~=~ \cos^2{\alpha\over 2}\; \density{0} 
           + \sin^2{\alpha\over 2}\; \density{a+1}.
\] 

In the second round, Bob replies with a random bit $b$.  So that she
wins, Alice sends~$a=b$ to Bob and subsequently tries to pass his
check. For that, she performs some unitary operation~$U_b$ on her part of the
state, and gets~$\ket{\tilde{\psi}_b}=(U_b\otimes I)\ket{\psi}$.
After that, she sends the part of the state in ${\mathcal H}_s$ (the
sign qubit) to Bob. The final joint state can be written now as
\[
\ket{\tilde{\psi}_b} ~=~ \sum_i\sqrt{p_i}\ket{i}\ket{\tilde{\psi}_{i,b}}.
\]
As we see, at the end of the protocol Bob has the density
matrix~$\sigma_b = \sum_i p_i
\ket{\tilde{\psi}_{i,b}}\bra{\tilde{\psi}_{i,b}}$.

The probability that Alice wins the game is equal to the probability
that she passes Bob's check at the end of the protocol, i.e.\ that Bob
measures his part of the joint state and gets~$\ket{\psi_b}$ as the
outcome
\begin{eqnarray*}
\Pr[\mbox{Alice wins }|\mbox{ Bob sends }b] 
  & = &    \sum_i p_i |\braket{\psi_b}{\tilde{\psi}_{i,b}}|^2 \\
  & = &    F(\sigma_b,\density{\psi_b}) \\
  & \leq & F(\trace_{{\mathcal H}_s}(\sigma_b), 
             \trace_{{\mathcal H}_s} \density{\psi_b} ) \\
  & = &    F(\sigma,\rho_b),
\end{eqnarray*}
where~$F(\cdot,\cdot)$ is the fidelity of two density matrices.  Here,
we have used the fact that the fidelity between two states can only
increase when we trace out a part of the states. Note also that the
state $\trace_{{\mathcal H}_s}(\sigma_b)$ is equal to $\sigma$, which
is independent of $b$.

Finally we have,
\begin{eqnarray*}
\Pr[\mbox{Alice wins}]
  & \leq & \frac{1}{2} [F(\sigma,\rho_0) + F(\sigma,\rho_1)] \\
  & \leq & \frac{1}{2}[1+\sqrt{F(\rho_0,\rho_1)}\;] \\
  & = &    {1 \over 2} (1 + \cos^2{\alpha\over 2}).
\end{eqnarray*} 
The second inequality is due to~\cite[Lemma~2]{SpekkensR02a},
\cite[Lemma~3.2]{NayakSh02}. Also~$F(\rho_0,\rho_1) =
\trnorm{\sqrt{\rho_0}\sqrt{\rho_1}}^2=\cos^4{\alpha\over 2}$. This completes
the proof.
\end{proof} 

Note that the analysis above is tight in the sense that Alice can
cheat with probability equal to~${1 \over 2} (1 + \cos^2{\alpha\over
  2})$. She does this by preparing the state~$\ket{\psi_0} +
\ket{\psi_1}$ (normalised) and sending the qutrit to Bob in the first
round. In the third round, she sends~$a = b$, and the sign qutrit from
the above state.

If Bob is the dishonest player, we can show the following bound. 
\begin{lemma}
\label{thm-bob}
If Alice is honest, then~$\Pr(c_A = 1) ~\le~ \left( \frac{1}{\sqrt{2}}
\cos^2{\alpha\over 2} + \sin^2 \frac{\alpha}{2} \right)^2$.
\end{lemma}
\begin{proof}
A cheating Bob tries to infer the value of the bit~$a$ that Alice
picked from the qutrit he receives in round~1 so that he can send~$b =
\bar{a}$. However, he has to minimize the disturbance caused to the
over all state~$\ket{\psi_a}$.  Suppose that Bob applies the unitary
transformation~$U$ on~$\aitch_t \tensor \aitch \tensor \complex^2$ to
the qutrit he receives from Alice, some ancillary qubits, and a qubit
reserved for his reply, and that:
\begin{equation}
\label{eqn-mapping}
U ~~:~~ \ket{i} \ket{\bar{0}} \ket{0} \mapsto \ket{\phi_{i,0}} \ket{0}
+ \ket{\phi_{i,1}} \ket{1}.
\end{equation}
He measures the last qubit, and sends that across in round~$2$. If the
XOR of the bit he sent and the one that Alice picked is~$1$ (i.e., $b =
\bar{a}$), in round~$4$ he sends one qutrit (the~$\aitch_t$ part) from
the above state across to Alice. 

Assuming that Alice had picked~$a$, Bob's probability of
winning is:\footnote{In the following, for a vector~$\ket{u} \in
{\mathcal H}$, and a vector~$\ket{v} \in {\mathcal H} \tensor
{\mathcal K}$, by~$\braket{u}{v}$ we will mean the projection of~$v$
under the operator~$\density{u}\tensor I$.}
\begin{eqnarray*}
\lefteqn{
\norm{ \bra{\psi_a} \frac{1}{\sqrt{2}} \left(  \ket{0} \left(
\cos{\alpha\over 2} \ket{\phi_{0,\bar{a}}} + \sin{\alpha\over 2}
\ket{\phi_{a+1,\bar{a}}} \right)  + \ket{1} \left( \cos{\alpha\over 2}
\ket{\phi_{0,\bar{a}}} - \sin{\alpha\over 2} \ket{\phi_{a+1,\bar{a}}}
\right) \right) }^2 } \\
    & = & \frac{1}{4} \norm{ \bra{\psi_{a,0}} \left( \cos{\alpha\over 2}
          \ket{\phi_{0,\bar{a}}} + \sin{\alpha\over 2}
          \ket{\phi_{a+1,\bar{a}}} \right) +
          \bra{\psi_{a,1}} \left( \cos{\alpha\over 2}
          \ket{\phi_{0,\bar{a}}} - \sin{\alpha\over 2}
          \ket{\phi_{a+1,\bar{a}}} \right) }^2  \\
    & = & \frac{1}{4} \left\| \cos^2 {\alpha\over 2}
          \braket{0}{\phi_{0,\bar{a}}} + \cos{\alpha\over 2}
          \sin{\alpha\over 2} \braket{0}{\phi_{a+1,\bar{a}}} +
          \cos{\alpha\over 2} \sin{\alpha\over 2}
          \braket{a+1}{\phi_{0,\bar{a}}} + \sin^2 {\alpha\over 2}
          \braket{a+1}{\phi_{a+1,\bar{a}}} \right. \\
    &   & \left. ~ + \cos^2 {\alpha\over 2} \braket{0}{\phi_{0,\bar{a}}} -
          \cos{\alpha\over 2} \sin{\alpha\over 2}
          \braket{0}{\phi_{a+1,\bar{a}}} - \cos{\alpha\over 2}
          \sin{\alpha\over 2} \braket{a+1}{\phi_{0,\bar{a}}} +
          \sin^2 {\alpha\over 2} \braket{a+1}{\phi_{a+1,\bar{a}}}
          \right\|^2 \\
    & = & \norm{ \cos^2 {\alpha\over 2} \braket{0}{\phi_{0,\bar{a}}} +
          \sin^2 {\alpha\over 2} \braket{a+1}{\phi_{a+1,\bar{a}}} }^2
          \\
    &\le& \left( \cos^2 {\alpha\over 2}
          \norm{\braket{0}{\phi_{0,\bar{a}}}} + \sin^2 {\alpha\over 2}
          \norm{ \braket{a+1}{\phi_{a+1,\bar{a}}} } \right)^2 \\
    &\le& \left( \cos^2 {\alpha\over 2} \norm{\phi_{0,\bar{a}}} +
          \sin^2 {\alpha\over 2} \norm{\phi_{a+1,\bar{a}}} \right)^2 \\
    &\le& \left( \cos^2 {\alpha\over 2} \norm{\phi_{0,\bar{a}}} +
          \sin^2 {\alpha\over 2} \right)^2.
\end{eqnarray*}
Now, consider~$\Pr[\textrm{Bob wins}]$, which is the average of the
above expression over~$a \in \set{0,1}$. This is maximised
when~$\norm{\phi_{0,0}} =
\norm{\phi_{0,1}} = 1/\sqrt{2}$ (recall from
equation~(\ref{eqn-mapping}) that~$\norm{\phi_{0,0}}^2 +
\norm{\phi_{0,1}}^2 = 1$). Thus, the probability of Bob winning is
bounded by
$$
\left( \frac{1}{\sqrt{2}} \cos^2 {\alpha\over 2}
+ \sin^2 {\alpha\over 2} \right)^2, 
$$ 
as claimed.
\end{proof}

There is a cheating strategy for Bob that achieves the above
probability of success. Bob can use the following transformation on
the qutrit he receives and an ancillary qubit:
\begin{eqnarray*}
\ket{0}\ket{0} 
  & \mapsto & \ket{0} \tensor \frac{1}{\sqrt{2}} (\ket{0} + \ket{1}),
  \text{ and} \\
\ket{x+1}\ket{0}
  & \mapsto & \ket{x+1} \ket{x}, \text{ for}~x \in \set{0,1}.
\end{eqnarray*}
He then measures the ancilla to get the bit~$b$ he is supposed to send
in the second round.

As we vary the parameter~$\alpha$ from 0 to $\pi$ Alice's cheating
probability decreases from~$1$ to~$1/2$ and Bob's cheating probability
increases from $1/2$ to~$1$. The bias is minimized when the two
probabilities are made equal:
\begin{eqnarray}
 {1 \over 2} (1 + \cos^2{\alpha\over 2})& = & \left( \frac{1}{\sqrt{2}}
\cos^2 {\alpha\over 2} 
+ \sin^2 {\alpha\over 2} \right)^2 
\end{eqnarray}  
By choosing $\alpha$ to satisfy the above equation, we get a protocol
in which no player can win the game with probability greater than
0.739. The bias is then~$0.239 < 1/4$.

\section{A strong coin flipping protocol}

Finally, we present a variant of the strong coin flipping protocol
of~\cite{Ambainis01}, which has the same bias, but is much more simple
to analyse.  The protocol has the following three rounds:
\begin{enumerate}
\item 
Alice picks~$a \inr \set{0,1}$, prepares the state~$\ket{\psi_{a}} \in
{\mathcal H}_s\otimes{\mathcal H}_t$ as in equation~(\ref{eqn-state})
and sends Bob the qutrit.

\item 
Bob picks~$b \inr \set{0,1}$ and sends it to Alice.
  
\item 
Alice then reveals the bit~$a$ to Bob and sends the second half of the
state~$\ket{\psi_a}$.  Bob checks that the qutrit-qubit pair he
received are indeed in state~$\ket{\psi_a}$.  If the test is passed,
Bob accepts the outcome $c=a\oplus b$, else Bob concludes that Alice
deviated from the protocol, and aborts.
\end{enumerate}

The analysis for Bob's cheating strategy is the same as
in \cite{Ambainis01} and his cheating probability is at most
\[ 
{1\over 2}+{\trnorm{\rho_0-\rho_1}\over 4} 
  ~=~ {1\over 2}(1+\sin^2 { \alpha \over 2}).
\] 

The analysis for Alice's cheating strategy is the same as in
Lemma~\ref{thm-alice} above, and the same bound of~$\frac{1}{2}(1 +
\cos^2 \frac{\alpha}{2})$ holds here as well. This analysis is
considerably simpler and does not require the symmetrization
in~\cite{Ambainis01} for the state sent in the first round.

By making the two cheating probabilities equal
\begin{eqnarray*}
{1 \over 2} (1 + \cos^2{\alpha\over 2}) 
  & = & {1\over 2}(1+\sin^2 { \alpha \over 2})
\end{eqnarray*} 
we achieve the bias of $1/4$ for $\alpha=\pi/2$.  

\subsection*{Acknowledgements}
We thank Umesh Vazirani for helpful discussions, and Rob Spekkens and
Terry Rudolph for detailed comments on the paper.

\end{document}